\documentstyle[epsfig,aps,pre,twocolumn,floats,eqsecnum]{revtex}

\begin{document}

\textheight= 240 truemm
\topmargin= -15 truemm

\newcommand{\gsim}{
\,\raisebox{0.35ex}{$>$}
\hspace{-1.7ex}\raisebox{-0.65ex}{$\sim$}\,
}

\newcommand{\lsim}{
\,\raisebox{0.35ex}{$<$}
\hspace{-1.7ex}\raisebox{-0.65ex}{$\sim$}\,
}

\bibliographystyle{prsty}

\title{
\begin{flushleft}
{\small 
PHYSICAL REVIEW E 
\hfill \qquad
VOLUME {\normalsize 54}, NUMBER {\normalsize 4}
\hfill 
OCTOBER {\normalsize 1996, 3250-3256}
}\\
\end{flushleft}  
Integral relaxation time of single-domain ferromagnetic particles
}

\author{
D.~A. Garanin
\renewcommand{\thefootnote}{\fnsymbol{footnote}}
\footnotemark[1]
}

\address{
I. Institut f\"ur Theoretische Physik, Universit\"at Hamburg,
Jungiusstrasse. 9, D-20355 Hamburg, Germany \\
%}
%\date{\today}
%\maketitle
%\abstract{
\smallskip
{\rm(Received 23 April 1996)}
\bigskip\\
\parbox{14.2cm}
{\rm
The integral relaxation time $\tau_{\rm int}$ of thermoactivating 
noninteracting 
single-domain ferromagnetic particles 
is calculated analytically in the geometry 
with a magnetic field $H$ applied parallel to the easy axis. 
It is shown that the drastic deviation of $\tau_{\rm int}^{-1}$ from the 
lowest 
eigenvalue of the Fokker-Planck equation $\Lambda_1$ at low 
temperatures, starting from some critical value of $H$, is the 
consequence of the depletion of the upper potential well.
In these conditions the integral relaxation time 
consists of two competing contributions corresponding to the overbarrier
and intrawell relaxation processes.
[S1063-651X(96)09409-3]
\smallskip
\begin{flushleft}
PACS number(s): 05.40.+j
\end{flushleft}
} 
} 
\maketitle

\renewcommand{\thefootnote}{\fnsymbol{footnote}}
\footnotetext[1]{ Electronic address: garanin@physnet.uni-hamburg.de }

\section{INTRODUCTION}

At present, a single-domain ferromagnetic particle with uniaxial anisotropy 
attracts the attention of researchers, in particular as one of 
the models of information storage.
The hysteretic rotation of the magnetization of such a particle over the 
potential barrier under the influence of an arbitrary directed magnetic 
field ${\bf H}$ was studied by 
Stoner and Wohlfart \cite{stowoh48}.
At nonzero temperatures the magnetization vector of the particle can 
surmount the barrier due to the thermal agitation, as argued by 
N\'{e}el \cite{nee49}; 
this effect becomes especially pronounced for small particles having 
lower values of the potential barrier $\Delta U$.
Such a ``superparamagnetic'' behavior was observed in many experiments 
on magnetic liquids, on polymers with magnetic inclusions, 
as well as on very thin magnetic layers forming ``islands.''

An initial accurate calculation of the thermoactivation rate of 
a uniaxial ferromagnetic particle is due to Brown \cite{bro63}, 
who derived the Fokker-Planck equation 
for an assembly of particles and solved it
in the presence of a longitudinal magnetic field 
${\bf H}=H{\bf e}_z$, 
perturbatively in the low-barrier case, $\Delta U \ll T$, 
and with the use of the Kramers transition-state method \cite{kra40} 
in the high-barrier limit $ T \ll \Delta U$ (the Boltzmann constant 
$k_B$ is set to unity).
In both limiting cases considered by Brown the time dependence of the 
average magnetization $\langle M_z\rangle$ is a single exponential
and the relaxation rate of ferromagnetic particles is given by the 
lowest eigenvalue $\Lambda_1$ of the Sturm-Liouville equation associated 
with the Fokker-Planck equation.
Subsequently, $\Lambda_1$ was calculated numerically by Aharoni for 
arbitrary values of $\Delta U/T$ without a magnetic field 
\cite{aha64} and with a longitudinal magnetic field \cite{aha69}.
The correction terms for the high-barrier result for $\Lambda_1$ were
given by Brown \cite{bro77}.
Later the analytical expression for $\Lambda_1$ in the high-barrier 
case was rederived in Ref. \cite{scucrecro92} with a more rigorous method. 
In Refs. \cite{besjafdor92,aha92,crecrowic94} 
various approximate analytical formulas 
for $\Lambda_1$ for the arbitrary $\Delta U/T$ were proposed.
Recently the thermoactivation rate of single-domain magnetic particles, 
as described by $\Lambda_1$, was calculated numerically by 
Coffey {\em et al.} \cite{cofetal95}
for the arbitrarily directed magnetic field ${\bf H}$, 
i.e., in the geometry considered by Stoner and Wohlfart \cite{stowoh48}.

Apart from limiting cases, the Fokker-Planck equation for an 
assembly of single-domain ferromagnetic particles cannot be solved 
analytically.
The magnetization relaxation curve consists of an infinite number of 
exponentials and the overall deviation 
of the linear dynamic susceptibility from the Debye form 
can be as large as about 7\% for 
{\em isotropic} particles in a static magnetic field, as shown in 
Ref. \cite{garishpan90}.
In this case it is convenient to introduce the so-called integral 
relaxation time $\tau_{\rm int}$, determined as the area 
under the relaxation 
curve after a sudden infinitesimal change of the magnetic field.
The quantity $\tau_{\rm int}$ depends on all eigenvalues $\Lambda_k$, 
$k=1,2,\ldots$, and is therefore more informative than $\Lambda_1$;  
also it can be directly measured.
Moreover, it turned out that, unlike $\Lambda_1$,
the integral relaxation time $\tau_{\rm int}$ 
can be calculated {\em analytically} for uniaxial particles 
in the longitudinal magnetic field for the arbitrary values 
of parameters \cite{garishpan90} 
and $\tau_{\rm int}^{-1}$ recovers the analytical results of Brown for 
$\Lambda_1$ in the asymptotic regions.

The integral relaxation time was 
also the subject of a recent series of papers 
\cite{cofetal94,walkalcof94,cofcrokalwal95}, where it was called 
the correlation time.
In Ref. \cite{cofetal94} $\tau_{\rm int}$ for 
uniaxial ferromagnetic particles was calculated {\em analytically} 
with an alternative method for zero magnetic field, the resulting 
expression being, however, much more complicated than the original 
formula for $\tau_{\rm int}$ of Ref. \cite{garishpan90}.
In Ref. \cite{cofcrokalwal95} a {\em numerical} calculation of 
$\tau_{\rm int}$ in the case with nonzero longitudinal magnetic 
field was presented.
In Ref. \cite{walkalcof94} the congeneric model of rotating dipoles 
describing the dielectric relaxation was considered.
The results of Ref. \cite{cofetal94} show that in zero magnetic field 
$\tau_{\rm int}^{-1}$ is very close to $\Lambda_1$ 
in the whole region of $\Delta U /T$. 
On the contrary, numerical calculations of Ref. \cite{cofcrokalwal95}
reveal a striking behavior $\tau_{\rm int}^{-1} \gg \Lambda_1$ for 
relatively small longitudinal fields in the region $ T \ll \Delta U$.
This region of parameters was not analyzed in Ref. \cite{garishpan90},
whereas in Ref. \cite{cofcrokalwal95} the effect was not physically 
interpreted.

The aim of this paper is thus to consider in more detail the integral 
relaxation time 
of uniaxial ferromagnetic particles in the longitudinal magnetic field 
with the help of the method of Ref. \cite{garishpan90}.
As we shall see, the effect found in Ref. \cite{cofcrokalwal95} can be 
explained by the depletion of the upper biased potential well, 
which leads to the dominance of the fast relaxation inside the 
lower well in the integral relaxation time.

The remainder of the paper is organized as follows.
In Sec. \ref{fopla} 
the derivation of 
the Fokker-Planck equation for an assembly of 
single-domain ferromagnetic particles from the stochastic 
Landau-Lifshitz equation is outlined. 
In Sec. \ref{lamtau}
the known results for the thermoactivation rate of uniaxial 
ferromagnetic particles are briefly reviewed.
Then the integral relaxation time $\tau_{\rm int}$ 
is introduced and analyzed, 
and it is shown that the effect discovered in Ref. \cite{cofcrokalwal95} 
can be explained without an explicit calculation of $\tau_{\rm int}$.
In Sec. \ref{taucalc}
the derivation of a general formula for $\tau_{\rm int}$ 
of uniaxial particles 
in a longitudinal magnetic field is 
presented and its behavior is studied analytically and numerically for 
the whole region of parameters.
Some concluding remarks are given in Sec. \ref{disc}.

\section{The Fokker-Planck equation}
\label{fopla}

The magnetization of a single-domain ferromagnetic particle ${\bf M}$ 
can be considered not too close to the Curie point $T_c$ 
as a vector of fixed length $|{\bf M}|=M_s(T)$, whose direction can 
fluctuate due to the thermal agitation.
This fluctuative motion of ${\bf M}$ can be described 
semiphenomenologically with the help of the stochastic equation 
%
%\marginpar{langeq}
%
\begin{equation}\label{langeq}
\dot {\bf M} = \gamma [{\bf M}\times 
( {\bf H}_{\rm eff} + \mbox{\boldmath$\zeta$} )] - {\bf R}({\bf M}) ,
\end{equation}
where $\gamma$ is the gyromagnetic ratio,
%
%\marginpar{heffw}
%
\begin{equation}\label{heffw}
{\bf H}_{\rm eff} = - \frac{\partial W}{\partial {\bf M}},
\qquad
W = -{\bf H\cdot M} - KM_z^2
\end{equation}
are the effective field and the energy density, ${\bf H}$ is the 
external magnetic field, and $K$ is the uniaxial anisotropy constant.
The energy of a particle is given by 
%
%\marginpar{hamvw}
%
\begin{equation}\label{hamvw}
{\cal H} = VW,
\end{equation}
where $V$ is the particle volume.
The correlators of different components of the white-noise field 
\mbox{\boldmath$\zeta$}$(t)$ can be conveniently written as
%
%\marginpar{zetacorr}
%
\begin{equation}\label{zetacorr}
\langle \zeta_i(t) \zeta_j(t') \rangle =
\frac{2\lambda T}{\gamma V} \delta_{ij} \delta(t-t') .
\end{equation}
The relaxation term ${\bf R}$ in (\ref{langeq}) describes, like 
\mbox{\boldmath$\zeta$}, the influence of the heat bath on the particle
and, as we shall see immediately, 
it has the Landau-Lifshitz form \cite{lanlif35}
%
%\marginpar{relterm}
%
\begin{equation}\label{relterm}
{\bf R} = \gamma\lambda [{\bf M}\times [{\bf M}\times {\bf H}_{\rm eff} ] ] .
\end{equation}

The Fokker-Planck equation corresponding to (\ref{langeq}) 
is formulated for the distribution function
$f({\bf N},t) = \langle \delta {\bf ( N} - {\bf M}(t){\bf )}\rangle$
on the sphere $|{\bf N}|=M_s$, where the average is taken over the 
realizations of \mbox{\boldmath$\zeta$}.
Differentiating $f$ over $t$ with the use of (\ref{langeq}) and 
calculating the right-hand side analogously to the derivations given, e.g., 
in Refs. \cite{mamaz75,zin89}, one comes to the Fokker-Planck equation 
%
%\marginpar{fpe}
%
\begin{eqnarray}\label{fpe}
&&
\frac{\partial f}{\partial t} =
- \frac{\partial}{\partial {\bf N}}
\bigg\{
\gamma \left [{\bf N}\times {\bf H}_{\rm eff} \right] 
- {\bf R}({\bf N})
\nonumber \\
&&
\qquad
+ \frac{\gamma\lambda T}{V} 
\left [{\bf N}\times 
\left [{\bf N}\times \frac{\partial}{\partial {\bf N}} 
\right ] 
\right ]
\bigg\} f .
\end{eqnarray}
One can easily see that the equilibrium distribution function
%
%\marginpar{fequi}
%
\begin{equation}\label{fequi}
f_0({\bf N}) \propto \exp[-{\cal H}({\bf N})/T],
\end{equation}
is the solution of (\ref{fpe}) if and only if ${\bf R}$ 
has the double-vector product form (\ref{relterm}),
which reflects the way of how magnetization is coupled to the 
heat-bath fluctuations in (\ref{langeq}).
If, e.g., the correlators of \mbox{\boldmath$\zeta$} components in 
(\ref{zetacorr}) are anisotropic, the expression (\ref{relterm}) also 
changes \cite{garishpan90}.

Brown used in his derivation of the Fokker-Planck equation \cite{bro63} 
the stochastic equation of motion (\ref{langeq}) 
with the Gilbert relaxation term
${\bf R} = \gamma\eta[{\bf M}\times \dot{\bf M}]$ \cite{gil55}.
Redefining $\gamma \Rightarrow \gamma_G$ in Brown's equation, one 
can transform the latter to the form (\ref{langeq}) and (\ref{relterm}) 
with $\gamma = \gamma_G/(1+\eta^2\gamma_G^2 M_s^2)$ and 
$\lambda = \eta \gamma_G$, where the 
Langevin field \mbox{\boldmath$\zeta$} enters also the expression for 
${\bf R}$ (\ref{relterm}) as being added to ${\bf H}_{\rm eff}$.
This means using a stochastic model somewhat different from the one 
described above.
Both models coinside, however, in the actual small-damping case 
$\eta\gamma M_s \ll 1$.

The equation of motion for the magnetization 
$\langle {\bf M}\rangle = \int d^3N {\bf N}f({\bf N},t)$
of an assembly of particles can be easily derived from (\ref{fpe}) 
and has the form
%
%\marginpar{fmomeq}
%
\begin{eqnarray}\label{fmomeq}
&&
\frac{d}{dt}\langle {\bf M}\rangle =
\gamma \langle [{\bf M}\times {\bf H}_{\rm eff}] \rangle
\nonumber \\
&&
\qquad
- \lambda \langle [{\bf M}\times [{\bf M}\times {\bf H}_{\rm eff}]]\rangle
- \Lambda_N \langle {\bf M}\rangle
\end{eqnarray}
[cf. (\ref{langeq})], where the characteristic diffusional relaxation 
rate $\Lambda_N$ is given by
%
%\marginpar{lamn}
%
\begin{equation}\label{lamn}
\Lambda_N \equiv \tau_N^{-1} \equiv 2\gamma\lambda T/V .
\end{equation}
One can see that even in the case without anisotropy, where
${\bf H}_{\rm eff}={\bf H}$, this equation is not closed since
it is connected 
to the second-order correlation functions $\langle M_i M_j \rangle$ in 
the Landau-Lifshitz term of (\ref{fmomeq}). 
Therefore, the resonance and relaxational behavior of the Fokker-Planck 
equation (\ref{fpe}) is in general {\em not} described by Lorentz and Debye 
curves, and the deviations from the latter can be about 7\% 
\cite{garishpan90}.
Neglecting these features, one can obtain the {\em best isolated} 
equation of motion 
for the magnetization of an assembly of particles in the isotropic 
case $K=0$, choosing the distribution function in the form 
$f({\bf N},t) \propto \exp[V {\bf A}(t){\bf N}/T]$
[cf. (\ref{fequi})],
where the temporal evolution of the vector ${\bf A}(t)$ 
is governed by (\ref{fmomeq}). 
Such a generalized Landau-Lifshitz-Bloch equation \cite{garishpan90}
contains both transverse and longitudinal relaxation terms.  
In the high-temperature limit (in the isotropic case 
$K=0$ this requires $T\gg VHM_s$) Eq. (\ref{fmomeq}) becomes closed 
and takes on the form of the Bloch equation with the relaxation rate 
$\Lambda_N$.

\section{
The lowest eigenvalue $\Lambda_1$ 
and the integral relaxation time $\tau_{\rm int}$ 
}
\label{lamtau}

To parametrize effects of thermal agitation on ferromagnetic 
particles, it is convenient to introduce the dimensionless energy 
$u\equiv {\cal H}/T = VW/T$, which in the case with a longitudinal 
magnetic field has the form
%
%\marginpar{uxial}
%
\begin{equation}\label{uxial}
u = -\xi x - \alpha x^2, \qquad x\equiv \cos\theta = M_z/M_s
\end{equation}
with
%
%\marginpar{xiald}
%
\begin{equation}\label{xiald}
\xi \equiv \frac{VHM_s}{T}, \qquad \alpha \equiv \frac{VKM_s^2}{T}.
\end{equation}
The top of the barrier corresponds to
%
%\marginpar{xm}
%
\begin{equation}\label{xm}
x = x_m = -h, \qquad h \equiv \frac{\xi}{2\alpha} = \frac{H}{2KM_s} .
\end{equation}
The barrier height $\Delta u \equiv u(x_m) - u(-1)$ is given by 
%
%\marginpar{delu}
%
\begin{equation}\label{delu}
\Delta u  = \alpha - \xi + \xi^2/(4\alpha) = \alpha (1-h)^2 .
\end{equation}

In the case $\alpha,\xi\sim 1$ a general solution of the 
Fokker-Planck equation (\ref{fpe}) cannot be found analytically and the 
relaxation of any initial state is described by a sum of exponentials
of the type 
$A_i \exp(-\Lambda_i t)$, where $\Lambda_i$ are the eigenvalues of  
the Sturm-Liouville equation associated 
with the Fokker-Planck equation 
(for the longitudinal relaxation all $\Lambda_i$ are real).
In the low-barrier case $\alpha,\xi\ll 1$, the problem can be solved 
perturbatively \cite{bro63} 
and the longitudinal relaxation is governed with a good 
accuracy by the single exponential corresponding to the lowest eigenvalue 
$\Lambda_1$ which is given by \cite{bro63}
%
%\marginpar{lam1sm}
%
\begin{equation}\label{lam1sm}
\Lambda_1 \cong \Lambda_N \left(1 - \frac{2}{5}\alpha + 
\frac{48}{875}\alpha^2 + \frac{1}{10}\xi^2 + \ldots \right) , 
\end{equation}
with $\Lambda_N$ determined by (\ref{lamn}).

In the high-barrier case $\alpha \gg 1$, the relaxation is dominated 
again by $\Lambda_1$ describing now the slow 
overbarrier thermoactivation, 
whereas all other eigenvalues $\Lambda_i$ correspond to the fast intrawell 
processes with small amplitudes.
Brown's result for the high-barrier case, which was derived with the 
help of the transition-state method of Kramers \cite{kra40}, 
can be written in the form \cite{aha69}
%
%\marginpar{lam1hi}
%
\begin{eqnarray}\label{lam1hi}
&&
\Lambda_1 \cong \Lambda_N \pi^{-1/2}\alpha^{3/2}(1-h^2)
 \\
&&
{}\times
\big\{ (1+h) \exp[-\alpha(1+h)^2] 
+ (1-h) \exp[-\alpha(1-h)^2] \big\},
\nonumber
\end{eqnarray}
where $h$ is given by (\ref{xm}).
The factor $(1+h)$ before the exponential function in (\ref{lam1hi}) is 
irrelevant since the first term of (\ref{lam1hi}) is only essential for 
$\xi\lsim 1$, which for $\alpha\gg 1$ implies $h\ll 1$.
We will, however, keep this factor here and in analogous expressions below 
for the sake of symmetry.

In the intermediate region $\alpha,\xi\sim 1$, it is convenient to 
introduce the integral relaxation time determined as the area under the 
magnetization relaxation curve 
after a sudden infinitesimal change of the applied field $H$ by 
$\Delta H$ at $t=0$:
%
%\marginpar{taudef}
%
\begin{equation}\label{taudef}
\tau_{\rm int} \equiv \int_0^\infty \!dt 
\frac
{\langle M_z(\infty)\rangle - \langle M_z(t)\rangle }
{\langle M_z(\infty)\rangle - \langle M_z(0)\rangle }.
\end{equation}
Unlike $\Lambda_1$, the integral relaxation time 
$\tau_{\rm int}$ can be found 
analytically from the Fokker-Planck equation (\ref{fpe}) 
in the whole range of 
parameters in the geometry with a longitudinal magnetic field 
\cite{garishpan90}, as will be described in detail in Sec. \ref{taucalc}.
Here we discuss the results of recent calculations of $\tau_{\rm int}$ by 
Coffey 
{\em et al.} \cite{cofetal94,cofcrokalwal95}.
At first note that the relaxation curve can be represented in the form 
%
%\marginpar{rcur}
%
\begin{equation}\label{rcur}
\langle M_z(\infty)\rangle - \langle M_z(t)\rangle
 = \Delta H \chi_z\sum_i A_i e^{-\Lambda_i t} ,
\end{equation}
where $\chi_z = \partial \langle M_z\rangle/\partial H$ 
is the static longitudinal susceptibility.
This form of writing the response function 
is more convenient than that of 
Refs. \cite{cofetal94,cofcrokalwal95} since here the amplitudes $A_i$ 
obey the sum rule $\sum_i A_i =1$. 
Now $\tau_{\rm int}$ of (\ref{taudef}) can be rewritten as (cf. 
\cite{cofetal94})
%
%\marginpar{taulam}
%
\begin{equation}\label{taulam}
\tau_{\rm int} = \sum_i A_i \Lambda_i^{-1} .
\end{equation}
In Refs. \cite{cofetal94,cofcrokalwal95} the integral relaxation time 
$\tau_{\rm int}$ is called the correlation time since according to the 
fluctuation-dissipation theorem $\tau_{\rm int}$ 
can be also considered as the area under the autocorrelation function.
The term ``correlation time,'' however, seems to be rather artificial 
because  
the autocorrelation function does not appear in the actual calculation 
of $\tau_{\rm int}$ with the help of (\ref{taudef}) or (\ref{taulam}), 
as well as in Sec. \ref{taucalc} below, and really  
considering autocorrelations would imply going unnecessarily 
beyond the Fokker-Planck equation.

According to the numerical results of Coffey {\em et al.} \cite{cofetal94}  
in a zero magnetic field the amplitudes $A_i$ satisfy 
$A_i \ll A_1$, $i=2,3,\ldots$, for all values of $\alpha$ 
and the difference between $\Lambda_1$ and $\tau_{\rm int}^{-1}$ is small 
everywhere reaching only 1.2\% at $\alpha=5$.
On the contrary, the subsequent calculations for $H\ne 0$ 
\cite{cofcrokalwal95} revealed a striking behavior 
$\tau_{\rm int}^{-1} \gg \Lambda_1$ at sufficiently low temperatures.
The formal reason for this is that $A_1$ becomes small in this region 
and the terms with $k=4,5$ dominate in (\ref{taulam}), as  
shown in Ref. \cite{cofcrokalwal95}.
But the effect can also be interpreted on a physical level as the 
consequence of the depletion of the upper potential well 
and quantitatively described without a general calculation of $\tau_{\rm 
int}$,
as will be demonstrated below.

The reduced equilibrium magnetization of an ensemble of 
noninteracting ferromagnetic particles
$m_z \equiv \langle M_z\rangle/M_s$
is given by the generalized Langevin function $B(\xi, \alpha)$:
%
%\marginpar{mavr}
%
\begin{equation}\label{mavr}
m_z = \int_{-1}^1 xf_0(x) dx = 
\frac{\partial}{\partial \xi} \ln Z = B(\xi, \alpha) , 
\end{equation}
where, according to (\ref{fequi}) and (\ref{uxial}),
%
%\marginpar{f0z}
%
\begin{equation}\label{f0z}
f_0 = \frac{1}{Z} \exp(-{\cal H}/T) = \frac{e^{-u}}{Z},
\qquad Z = \int_{-1}^1 e^{-u} dx .
\end{equation}
In the high-barrier case $\alpha \gg 1$, the partition function $Z$ is 
a sum of two contributions corresponding to the two potential wells
$Z = Z_+ + Z_-$,
%
%\marginpar{zalhi}
%
\begin{equation}\label{zalhi}
Z_\pm \cong \frac{e^{\alpha\pm\xi}}{2\alpha\pm\xi}
\left[ 1 + \frac{2\alpha}{(2\alpha\pm\xi)^2} + \ldots \right] , 
\end{equation}
where the correction terms account for the curvature of the 
potential-energy function $u$ of (\ref{uxial}).
Neglecting these small terms, one can represent 
$B(\xi,\alpha)$ of Eq. (\ref{mavr}) by two mutually complementing 
expressions
%
%\marginpar{bwb1}
%
\begin{equation}\label{bwb1}
B(\xi, \alpha) \cong 
\tanh\xi 
- \frac{1}{2\alpha} \left( \frac{\xi}{\cosh^2\xi} + \tanh\xi \right)
+ \frac{\xi}{(2\alpha)^2}
\end{equation}
for $1\sim\xi\ll\alpha$ and
%
%\marginpar{bwb2}
%
\begin{equation}\label{bwb2}
B(\xi, \alpha) \cong
1
- 2e^{-2\xi} \frac{2\alpha+\xi}{2\alpha-\xi}
- \frac{1}{2\alpha+\xi}      
\end{equation}
for  $1\ll\xi\sim\alpha$.
Here, in the first limiting expression the second term 
is small and irrelevant; the third term is kept since it yields a 
contribution to the derivative $B' = \partial B/\partial \xi$ that is 
not exponentially small for $\xi\gg 1$.
In the second limiting expression (the strong-bias case) the deviation
of $B$ from unity separates into two parts: 
The second exponentially small term is due to the 
population of the upper well ($x\sim -1$), whereas the third one 
accounts for the thermal agitation in the lower well ($x\sim 1$).
The response of magnetization to an infinitesimal change of the 
magnetic field $\Delta H$ 
is related to the derivative of the generalized Langevin function $B'$:
$\Delta m = B'\Delta H/T$.
The latter can be determined from (\ref{bwb1}) and (\ref{bwb2}) and put 
in the whole region $2\alpha-\xi \gg 1$
into the unique expression 
%
%\marginpar{bdwb}
%
\begin{eqnarray}\label{bdwb}
&&
B' \cong B'_B + B'_W \cong
\frac{1}{\cosh^2\xi}\frac{2\alpha+\xi}{2\alpha-\xi}
+ \frac{1}{(2\alpha+\xi)^2} \nonumber \\
&&\qquad 
\cong \frac{1-h^2}{c^2(\xi,h)} + \frac{1}{(2\alpha+\xi)^2} , 
\end{eqnarray}
where 
%
%\marginpar{cxih}
%
\begin{equation}\label{cxih}
c(\xi,h) \equiv \frac{1}{2} \big[ (1+h)e^{-\xi} + (1-h)e^\xi \big] 
\end{equation}
and $B'_B$ accounts for the redistribution of particles between the two 
wells across the potential barrier and $B'_W$ that inside the lower 
well.
Henceforth we will use the second of the equivalent forms of $B'$ in
(\ref{bdwb}) for the sake of symmetry 
[cf. the comments after Eq. (\ref{lam1hi})].
\begin{figure}[t]
\unitlength1cm
\begin{picture}(11,7)
\centerline{\epsfig{file=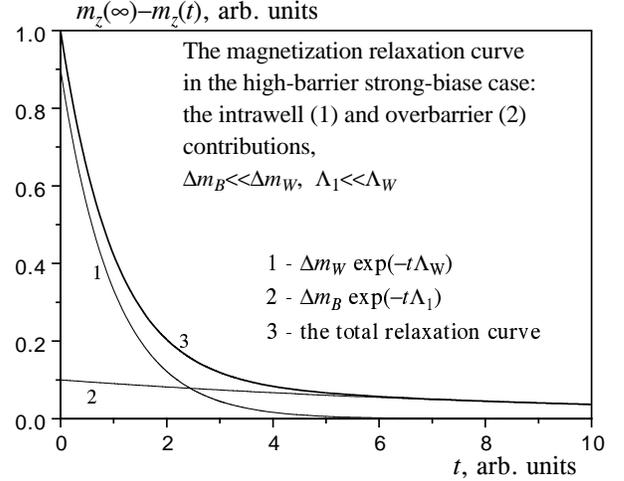,angle=-90,width=12cm}}
\end{picture}
\caption{ \label{tau_rel} 
A schematic look of the two-exponential relaxation curve 
of single-domain ferromagnetic 
particles in the strong-biase high-barrier case, $\alpha,\xi\gg1$. 
}
\end{figure}

Now, in the low-temperature strong-bias case $\alpha,\xi\gg1$, 
the relaxation curve (\ref{rcur}) consists of only two exponentials 
(see Fig. \ref{tau_rel})
%
%\marginpar{delmt}
%
\begin{eqnarray}\label{delmt}
&&
m_z(\infty) - m_z(t)  
\nonumber \\
&&\qquad
= \Delta m_B\exp(-t\Lambda_1) + \Delta m_W\exp(-t\Lambda_W) ,
\end{eqnarray}
where $\Lambda_1$ is given by (\ref{lam1hi}) and
%
%\marginpar{tauw}
%
\begin{equation}\label{tauw}
\Lambda_W\cong 2\gamma\lambda H_{\rm eff} = \Lambda_N (2\alpha+\xi)
\end{equation}
is the temperature-independent relaxation rate in the lower well, which 
can be obtained from the deterministic Landau-Lifshitz equation 
(\ref{langeq}) and (\ref{relterm}) without $\bbox{\zeta}$. 
The integral relaxation time $\tau_{\rm int}$ calculated according to the 
definition (\ref{taudef}) can be written as
%
%\marginpar{tauwb0}
%
\begin{equation}\label{tauwb0}
\tau_{\rm int} \cong \tau_{{\rm int},B} + \tau_{{\rm int},W}, 
\end{equation}
where
%
%\marginpar{tauwb}
%
\begin{equation}\label{tauwb}
\tau_{{\rm int},B} = \frac{B'_B}{B'}\Lambda_1^{-1}, 
\qquad \tau_{{\rm int},W} = \frac{B'_W}{B'}\Lambda_W^{-1} .
\end{equation}
One can see that 
in the low-temperature strong-bias case $\alpha,\xi\gg1$, 
the barrier contribution $\tau_{{\rm int},B}$ into the integral 
relaxation time 
$\tau_{\rm int}$ can be substantially reduced due to the depletion of 
the upper 
potential well manifesting itself in the exponential smallness 
of the magnetization change due to overbarrier transitions 
$\Delta m_B\propto B'_B$ [see (\ref{bdwb})].
In this case the overbarrier and intrawell terms in (\ref{tauwb0}) can 
compete with each other since $\Lambda_1$ is exponentially small and 
$B'_W/B'\cong 1$.
On the contrary, for small or zero bias one has $B'_B/B'\cong 1$ and 
$B'_W/B'\ll 1$, so that the intrawell process can be completely 
ignored.

The expressions for $\tau_{{\rm int},B}$ and $\tau_{{\rm int},W}$ in 
(\ref{tauwb}) are valid in 
the whole high-barrier region $2\alpha-\xi \gg 1$ and will be 
obtained independently in the framework of a general method in 
Sec. \ref{taucalc}.
In the strong-bias case $\xi\gg 1$, 
the barrier contribution $\tau_{{\rm int},B}$ in (\ref{tauwb}) can be 
represented
with the use of (\ref{lam1hi}), (\ref{bdwb}), and (\ref{xm}) as
%
%\marginpar{taubhi}
%
\begin{equation}\label{taubhi}
\tau_{{\rm int},B} = 16 \Lambda_N^{-1}(\pi\alpha)^{1/2}\frac{(1+h)^2}{(1-h)^3}
\exp[\alpha(1-6h+h^2)] .
\end{equation}
It changes its behavior as a function of $\alpha$ at the critical value
of the applied field
%
%\marginpar{hc}
%
\begin{equation}\label{hc}
h = h_c = 3-2\sqrt{2} \approx 0.17 ,
\end{equation}
which is substantially smaller than the field of the barrier 
disappearance $h=1$ [see (\ref{xm}) and (\ref{delu})].
For $h$ in the vicinity of $h_c$ the exponential factor in (\ref{taubhi})
can be written as $\exp[-4\sqrt{2}\alpha(h-h_c)]$.
It can be seen that for $h<h_c$ the quantity $\tau_{{\rm int},B}$ 
exponentially 
increases with lowering 
temperature (i.e., with increasing of $\alpha$) 
and brings the dominant contribution into $\tau_{\rm int}$ of Eq. 
(\ref{tauwb}).
On the contrary, for $h>h_c$ the quantity $\tau_{{\rm int},B}$ exponentially 
decreases at large $\alpha$, so that $\tau_{\rm int}$ tends to the 
temperature-independent value $\tau_{{\rm int},W}$ of (\ref{tauw}).
One can also see that for $h$ only slightly higher than $h_c$ the quantity 
$\tau_{{\rm int},B}$ increases as $\alpha^{1/2}$ at smaller $\alpha$; 
then the decreasing exponential becomes to dominate.
Thus, in this case $\tau_{{\rm int},B}$, and hence $\tau_{\rm int}$ of 
(\ref{tauwb}), has a 
maximum at some $\alpha\gg 1$ and $\tau_{\rm int}^{-1}$ 
has the corresponding minimum, as was obtained numerically in 
Ref. \cite{cofcrokalwal95}.
It should be noted, however, that the actual position of this minimum  
can be described only taking into account in (\ref{tauwb}) 
the general form of $B'$ given by (\ref{bdwb}).

The results above completely describe the observations made in 
Ref. \cite{cofcrokalwal95} in the low-temperature strong-bias region.
In the next section we present the analytical calculation of $\tau_{\rm 
int}$ in 
the whole range of parameters.

\section{Calculation of the integral relaxation time $\tau_{\rm int}$}
\label{taucalc}

All the information about the relaxation curve (\ref{rcur}) is 
contained in the longitudinal linear 
dynamic susceptibility $\chi(\omega)$.
In the presence of a small alternating field 
$\Delta H_z(t) = \Delta H_{z0}\exp(-i\omega t)$ 
the deviation of the distribution function $f$ from the equilibrium function 
(\ref{f0z}) can be represented as
%
%\marginpar{df}
%
\begin{equation}\label{df}
\delta f = f_0(x) q(x) VM_s\Delta H_z(t)/T,
\end{equation}
where the function $q(x)$ satisfies an equation following from 
(\ref{fpe}):
%
%\marginpar{qeq}
%
\begin{eqnarray}\label{qeq}
&&
\left( \frac{d}{dx} + 2\alpha x + \xi \right)
(1-x^2)\frac{dq}{dx} + 2i\omega \Lambda_N^{-1}q 
\nonumber \\
&&\qquad\qquad
= (1-x^2)(2\alpha x + \xi) - 2x .
\end{eqnarray}
The dynamic susceptibility of the particle's assembly 
is then determined by
%
%\marginpar{dsus}
%
\begin{equation}\label{dsus}
\chi_z(\omega) = \frac{VM_s^2}{T}\int_{-1}^1 \! dx\,x f_0(x) q(x) .
\end{equation}

Using the linear-response theory one can easily show that  
$\chi_z(\omega)$ can be represented in the form
%
%\marginpar{chialam}
%
\begin{equation}\label{chialam}
\chi_z(\omega) = \chi_z \sum_i \frac{A_i}{1-i\omega\Lambda_i^{-1}} ,
\end{equation}
where $\chi_z$, $A_i$, and $\Lambda_i$ are the parameters of the 
magnetization relaxation curve (\ref{rcur}).
Calculating this sum of Debye terms requires knowing all eigenvalues 
$\Lambda_i$ and amplitudes $A_i$ associated with the Fokker-Planck 
equation and cannot be done analytically in the general case.
Accordingly, Eq. (\ref{qeq}) has no general analytical solution 
and its behavior is to be studied analytically in the limiting cases of 
high and low temperatures and high and low frequencies, as was
done in Ref. \cite{garishpan90}.
In particular, generating high-frequency expansions of $\chi(\omega)$
does not require solving differrential equations and can be carried out 
up to high orders.
The corresponding results, however, are not very interesting here 
since they describe only fast intrawell processes.
The information about the slow process of thermoactivation is contained 
in the low-frequency expansion of $\chi(\omega)$, which can be written 
in the form
%
%\marginpar{sustau}
%
\begin{equation}\label{sustau}
\chi_z(\omega) \cong \chi_z (1 + i\omega\tau_{\rm int} + \ldots) .
\end{equation}
Comparing (\ref{sustau}) with (\ref{chialam}), one can show 
that the quantity $\tau_{\rm int}$ in (\ref{sustau}) is exactly the 
integral relaxation time given by the formula (\ref{taulam}).

The perturbative solution of (\ref{qeq}) for small $\omega$ 
can be done analytically since for $\omega=0$ there are only terms of 
the type $q'(x)$ and $q''(x)$ in the equation.
Hence one can introduce a new variable $g(x)\equiv q'(x)$ and solve 
successively the first-order differential equations for $g(x)$ and 
$q(x)$.
After calculation of the susceptibility (\ref{dsus}) one gets the 
analytic expression the for integral relaxation time 
$\tau_{\rm int}$ \cite{garishpan90}:
%
%\marginpar{tauq}
%
\begin{equation}\label{tauq}
\tau_{\rm int} = \frac{2}{\Lambda_N B'} 
\int_{-1}^1 \frac{dx}{1-x^2}\Phi^2(x)f_0^{-1}(x) ,
\end{equation}
where $f_0$ is given by (\ref{f0z}), $B' = \partial B/\partial \xi$, and 
%
%\marginpar{phix}
%
\begin{equation}\label{phix}
\Phi(x) = \int_{-1}^x (B - x') f_0(x') dx' .
\end{equation}
Recalling the general formula for $B(\xi,\alpha)$ Eq. (\ref{mavr}), 
one can conclude that $\Phi(\pm 1)=0$, i.e., 
the integrand of (\ref{tauq}) goes to zero at $x=\pm 1$.
The function $\Phi(x)$ can be easily calculated analytically 
in two particular cases.
In the unbiased case $\xi=0$ one gets
%
%\marginpar{phial}
%
\begin{equation}\label{phial}
\Phi(x) = \frac{1}{2\alpha} [ f_0(1) - f_0(x) ],
\qquad f_0(x)=\frac{ \exp(\alpha x^2) }{ Z(\alpha) },
\end{equation}
whereas in the isotropic case $\alpha=0$ 
%
%\marginpar{phixi}
%
\begin{equation}\label{phixi}
\Phi(x) = \frac{f_0(x)}{\xi}
\left[ \coth\xi - x - \frac{\exp(-\xi x)}{\sinh\xi} \right] , 
\end{equation}
with $f_0(x)=\xi\exp(\xi x)/(2\sinh\xi)$.
An alternative analytical expression for $\tau_{\rm int}$ in the 
particular case 
$\xi=0$ was derived recently by Coffey {\em et al.} \cite{cofetal94} with 
the help of the development of the solution of the Fokker-Planck 
equation in Legendre polynomials.
Their expression contains Kummer functions and is essentially more 
complicated than (\ref{tauq}) with (\ref{phial}).

Now we proceed with the analysis of the general expression (\ref{tauq}) 
in different limiting cases.
In the high-temperature limit $\alpha,\xi \ll 1$, the calculation can 
be done perturbatively with respect to $\alpha$ and $\xi$.
In particular, at very large temperatures one has 
$f_0(x)\cong 1/2$, $B\cong0$, $B'\cong1/3$, and $\Phi\cong(1-x^2)/4$.
Thus the whole phase space of the ferromagnetic particle, 
$-1\leq x \leq 1$, contributes to (\ref{tauq}) 
and one gets $\tau_{\rm int}^{-1}\cong \Lambda_N$.
A more accurate calculation yields
%
%\marginpar{tauht}
%
\begin{equation}\label{tauht}
\tau_{\rm int}^{-1}\cong \Lambda_N 
\left(1 - \frac{2}{5}\alpha + 
\frac{2}{35}\alpha^2 + \frac{1}{9}\xi^2 + \ldots \right) ,
\end{equation}
which is very close to Brown's expression for $\Lambda_1$ given by 
(\ref{lam1sm}).

In the unbiased low-temperature case $\xi=0$, $\alpha\gg1$, the 
function $\Phi$ given by (\ref{phial}) is constant in the main part of 
the $x$ interval, except for near the borders.
Thus the main contribution to the integral (\ref{tauq}) comes from the 
barrier region $x\sim\alpha^{-1/2} \ll 1$ cut by the function 
$f_0^{-1}\propto \exp(-\alpha x^2)$.
With the use of (\ref{zalhi}) with $\xi=0$ one gets, in the leading order,
%
%\marginpar{taual}
%
\begin{equation}\label{taual}
\tau_{\rm int}^{-1}\cong 2\Lambda_N \pi^{-1/2}\alpha^{3/2}e^{-\alpha} ,
\end{equation}
which coincides with the expression for $\Lambda_1$ in (\ref{lam1hi}) in 
the unbiased case $h=0$.
It is not difficult to calculate also the correction terms for the 
formula (\ref{taual}), which coincide with those given by Brown 
\cite{bro77}.
In the isotropic strong-field limit $\alpha=0$, $\xi\gg 1$, the 
function $\Phi$ given by (\ref{phixi}), as well as the whole integrand of 
(\ref{tauq}), is peaked in the vicinity of the potential minimum $x=1$, 
where the exponentially small term with $\exp(-\xi x)$ can be neglected.
In the leading order one gets for $\tau_{\rm int}^{-1}$ 
the temperature-independent expression (\ref{tauw}) with $\alpha=0$. 
\begin{figure}[t]
\unitlength1cm
\begin{picture}(11,7)
\centerline{\epsfig{file=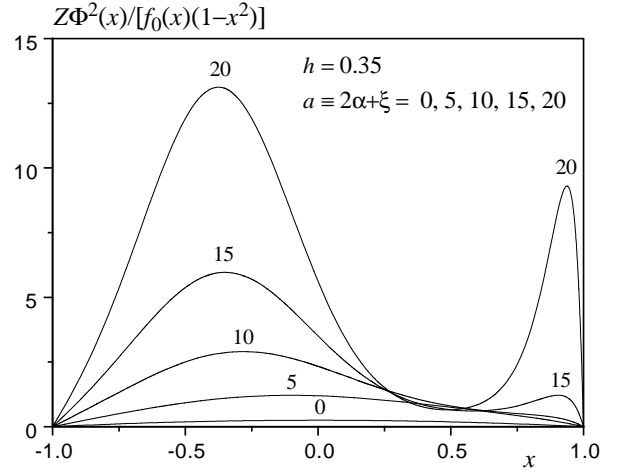,angle=-90,width=12cm}}
\end{picture}
\caption{ \label{tau_test} 
The integrand of the formula (\protect\ref{tauq}) 
for the integral relaxation time $\tau_{\rm int}$ for different values of 
parameters.
}
\end{figure}

Now we consider, as a corollary, the biased low-temperature case 
$\xi\ne0$, $\alpha\gg1$.
As argued in Sec. \ref{lamtau}, in this case $\tau_{\rm int}$ can be 
the sum 
of intrawell and overbarrier contributions $\tau_{{\rm int},W}$ and 
$\tau_{{\rm int},B}$ 
[see (\ref{tauwb0})].
Accordingly, the integrand of (\ref{tauq}) can consist, 
for $\alpha,\xi\gg 1$, 
of two peaks (see Fig. \ref{tau_test}) corresponding to the 
barrier top $x\sim x_m=-h$ in (\ref{xm}) and the lower well 
$x\sim 1$.
The function $\Phi(x)$ of (\ref{phix}) is determined by two 
well-separated potential wells and is therefore practically 
independent of $x$ for $x$ not too close to the boundaries.
In the the lower-well region $1-x \ll 1$, the calculation yields  
$\Phi(x)$ as a sum of two contributions 
$\Phi = \Phi_B + \Phi_W$, where to the leading order
%
%\marginpar{phib}
%
\begin{equation}\label{phib}
\Phi_B(x) \cong 
\frac{1-h^2}{2c^2(\xi,h)}
\big\{ 1 - \exp[ -(2\alpha+\xi)(1-x) ] \big\} ,
\end{equation}
$c(\xi,h)$ is given by (\ref{cxih}) and 
%
%\marginpar{phiw}
%
\begin{equation}\label{phiw}
\Phi_W(x) \cong (1-x) \exp[ -(2\alpha+\xi)(1-x) ] .
\end{equation}
The term $\Phi_B$ of (\ref{phib}) goes over to the constant mentioned 
above in the region not too close to the border 
[$1-x \gg 1/(2\alpha+\xi)\ll 1$]. 
For $\xi\gg 1$ it acquires the small factor $e^{-2\xi}$ 
accounting for the depletion of the upper potential well 
[cf. (\ref{bdwb}) and (\ref{tauwb})].
On the contrary, such a factor is not present in $\Phi_W$ in 
(\ref{phiw}), but the corresponding contribution into $\tau_{\rm int}$ 
given by 
(\ref{tauq}) is reduced due to $f_0^{-1}(x)$.
Now calculating the integral (\ref{tauq}) one gets Eq. (\ref{tauwb0}),
where
%
%\marginpar{taub}
%
\begin{equation}\label{taub}
\tau_{{\rm int},B} \cong \frac{B'_B}{B'}
\frac{\pi^{1/2}}{2\Lambda_N \alpha^{3/2}}
\frac{\exp[\alpha+\xi^2/(4\alpha)]}
{(1-h^2)c(\xi,h)}
\end{equation}
coincides with the expression given by (\ref{tauwb}) and in the 
strong-bias case
%
%\marginpar{tauw1}
%
\begin{equation}\label{tauw1}
\tau_{{\rm int},W}^{-1} \cong \Lambda_W \cong \Lambda_N 
\left( 2\alpha+\xi - \frac{10\alpha+\xi}{2\alpha+\xi} \right) .
\end{equation}
In (\ref{tauw1}) [cf. (\ref{tauw})] the correction terms, 
in particular of the type present in (\ref{zalhi}), have been taken into 
account.
\begin{figure}[t]
\unitlength1cm
\begin{picture}(11,7)
\centerline{\epsfig{file=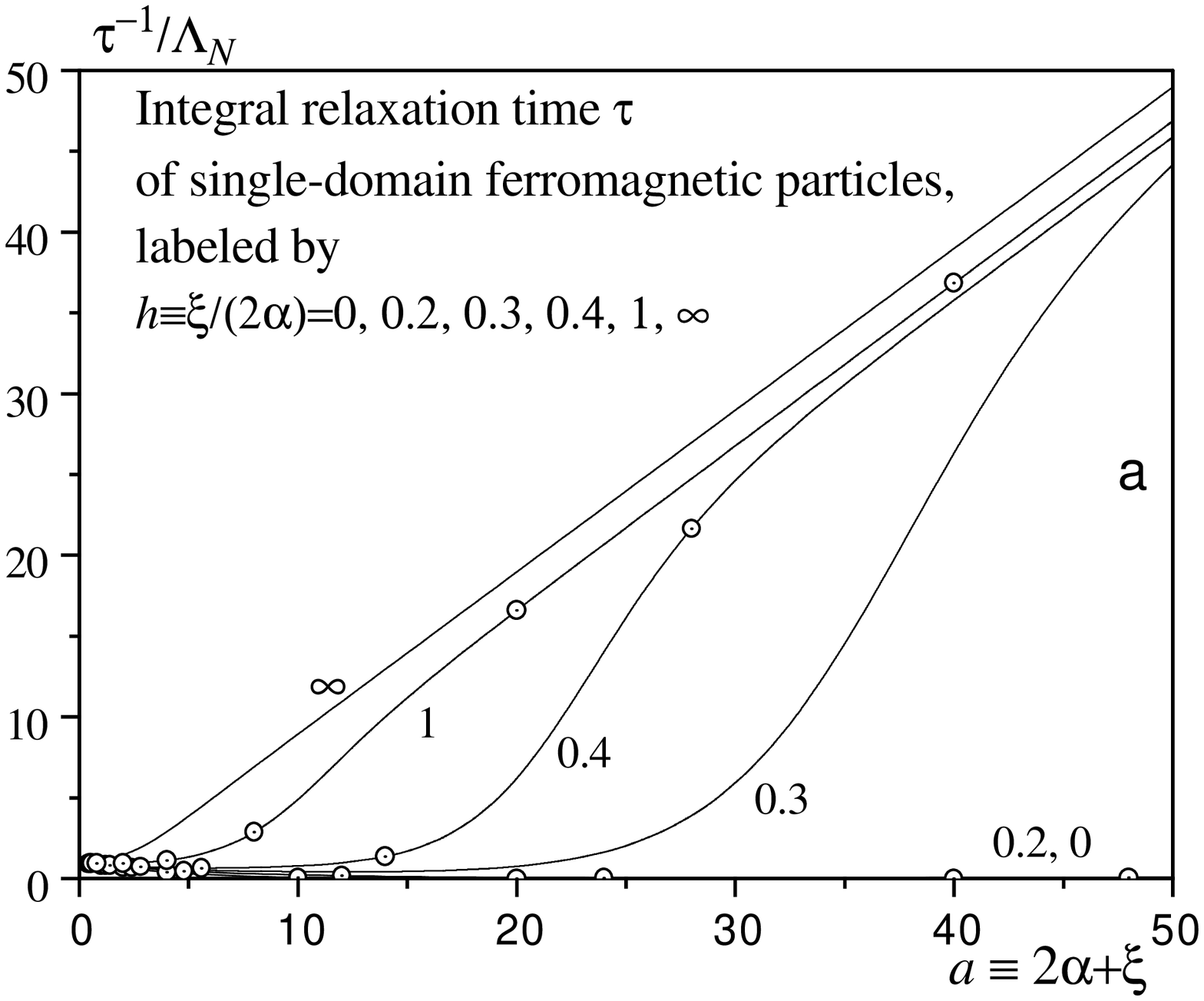,angle=-90,width=12cm}}
\end{picture}
\begin{picture}(11,6.5)
\centerline{\epsfig{file=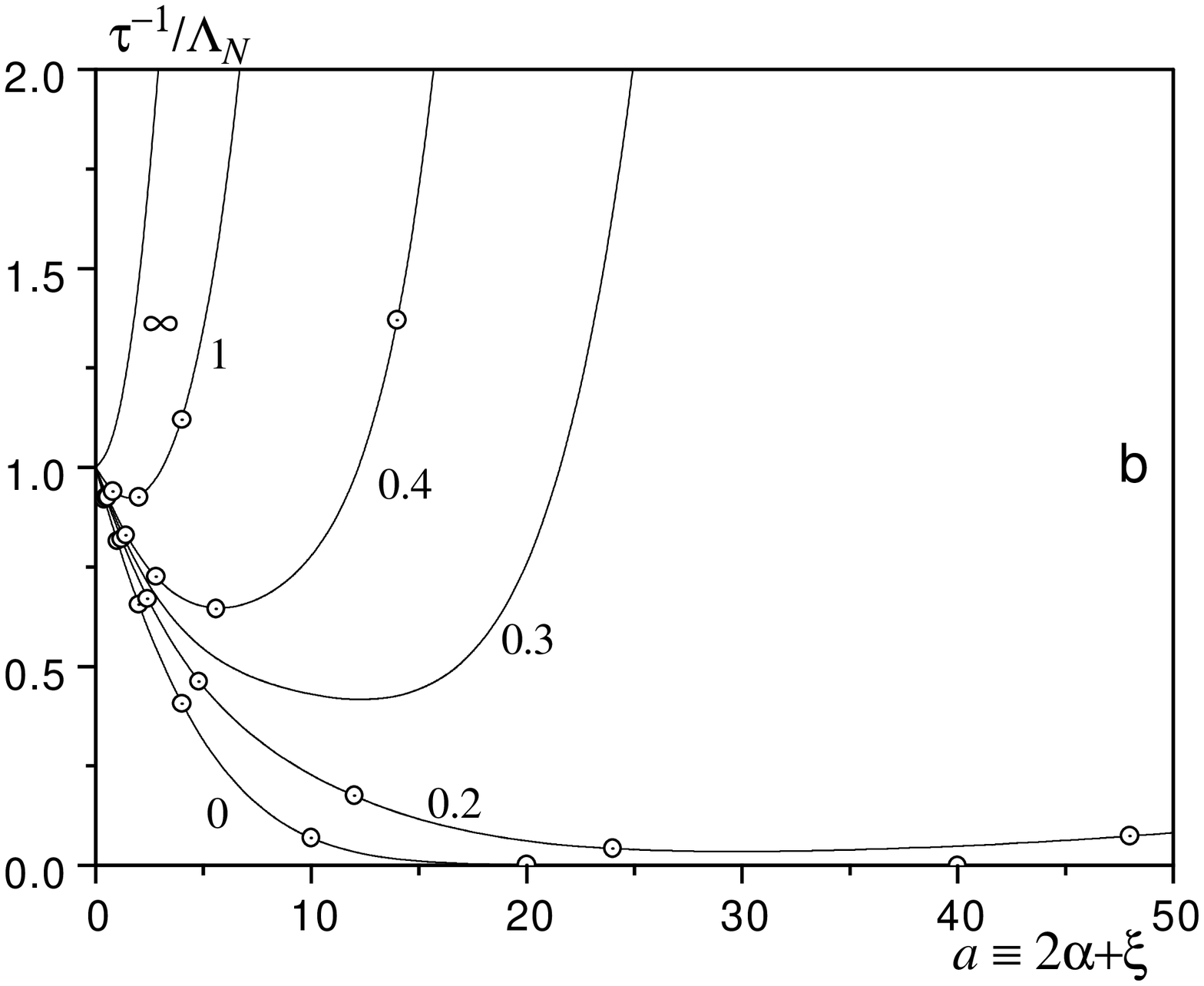,angle=-90,width=12cm}}
\end{picture}
\caption{ \label{tau_res} 
Inverse of the integral relaxation time $\tau_{\rm int}$ as a function of the 
parameter $a$ of (\protect\ref{adef}). Circles: data of 
Coffey {\em et al.} \protect\cite{cofcrokalwal95} taken from their Tab. II. 
}
\end{figure}
\par
Numerical calculation of the integral relaxation time $\tau_{\rm int}$ 
given by 
(\ref{tauq}) in the whole range of parameters $\alpha$ and $\xi$ poses 
no difficulties.
For the representation of the results 
for the arbitrary relation between $\alpha$ and $\xi$, including the 
case $\alpha=0$, it is more convenient to use the variable
%
%\marginpar{adef}
%
\begin{equation}\label{adef}
a \equiv 2\alpha + \xi = \frac{VM_s}{\gamma T} \omega_R,
\end{equation}
where $\omega_R = \gamma (2KM_s+H)$ is the ferromagnetic
resonance frequency 
in the lower potential well.
One can see that $2\alpha = a/(1+h)$ and 
$\xi=ah/(1+h)$.
In terms of the variables $a$ and $h$ the asymptotic formula 
(\ref{tauw1}) can be rewritten as
$\tau_{{\rm int},W}^{-1} \cong \Lambda_N [ a - (5+h)/(1+h) ]$,  
which shows that all curves $\tau_{{\rm int},W}^{-1}(a)$ for different 
$h$ are parallel to each other.
The results of the numerical integration in the formula (\ref{tauq}) are 
represented in Fig. \ref{tau_res}.
These results confirm all the considerations made above, as well as the 
numerical findings of Coffey {\em et al.} \cite{cofcrokalwal95}.

\section{Discussion}
\label{disc}

The formula for the integral relaxation time $\tau_{\rm int}$ of an 
assembly of 
single-domain ferromagnetic particles (\ref{tauq}) 
is the all-temperature solution of 
the problem of the thermoactivation escape rate of the particles over a 
potential barrier.
In the unbiased case $\xi = h = 0$, the quantity $\tau_{\rm int}$ is 
very close 
in the whole temperature range 
to the inverse of the lowest eigenvalue of the 
Fokker-Planck equation $\Lambda_1$.
Besides, in comparison to the latter, 
$\tau_{\rm int}$ has more physical significance and is given by a simple 
quadrature (\ref{tauq}) with (\ref{phial}).
In the high-barrier strong-bias case $\alpha, \xi \gg1$, as a result of 
the depletion of the upper potential well the relaxation is described by 
two exponentials with uncomparable characteristic times, which 
correspond to the intrawell and overbarrier processes.
In this case it would be naturally an oversimplification to describe the 
relaxation with a common integral relaxation time, but it 
is possible to separate analytically both contributions in the general 
formula (\ref{tauq}) [see (\ref{taub}) and (\ref{tauw1})] and thus to 
give a complete description of the relaxation process.

Going beyond the integral relaxation time, one can conceive the 
calculation of the whole dynamic susceptibility $\chi_z(\omega)$ given 
by the formula (\ref{chialam}).
In this case the numerical approach of Ref. \cite{cofcrokalwal95}, 
consisting of the calculation of all $A_i$ and $\Lambda_i$, is really 
useful.
However, for the model with a high potential barrier 
such a calculation is 
probably not very interesting for the reasons mentioned above: there are, 
to a good accuracy, only one (in the unbiased case) 
or two (in the strong-bias case) terms in (\ref{chialam}).
More appealing would be to produce calculations for an isotropic model 
in a field ($\alpha=0$, $\xi\ne0$) where the eigenvalues $\Lambda_i$ are 
not so well separated.
In this case deviations from the Debye form of the longitudinal dynamic 
susceptibility were studied analytically in Ref. \cite{garishpan90}, 
where it was shown that these deviations can be about 7\% at 
$\xi\approx 3$.
In Ref. \cite{garishpan90} an effective two-relaxator 
formula for $\chi_z(\omega)$ was proposed, 
which was argued to describe the main part of deviations from the simple 
Debye form.
It would be interesting to check and improve these results by a direct 
numerical calculation.

One more unsolved problem is the calculation of the 
{\em transverse} integral relaxation time of superparamagnetic 
particles. 
Up to now it seems to be considered only for the model 
of rotating dipoles in Ref. \cite{walkalcof94}.

Although many experimental investigations are currently done on systems 
showing superparamagnetism, these investigations are practically 
confined to the certification of a superparamagnetic behavior and to 
rough estimation of relaxation times.
It would be worth making more purposeful measurements aimed at a 
comparison with existing theories.
For this purpose it would be important to eliminate the distribution of  
particle volumes and orientations of the anisotropy axes.

\section*{Acknowledgments}

The author thanks Hartwig Schmidt for valuable discussions.
The financial support of Deutsche Forschungsgemeinschaft 
under contract No. Schm 398/5-1 is greatfully acknowledged.

\vspace{-0.5cm}

%\bibliography{gar}

\end{document}